\documentclass[aps,prb,showpacs,twocolumn,superscriptaddress]{revtex4-2}
\usepackage{amssymb}
\usepackage{graphicx}
\usepackage{appendix}
\usepackage{dcolumn}
\usepackage{bm}
\usepackage{amsmath}
\usepackage{epstopdf}
\usepackage{xcolor}
\usepackage{float}
\usepackage{braket}
\usepackage[colorlinks,urlcolor=blue,anchorcolor=blue,linkcolor=blue,citecolor=blue,breaklinks=true]{hyperref}
\usepackage[T1]{fontenc}
\usepackage{array}
\usepackage{booktabs}

\begin{document}
\title{Alternative $\nu+\nu$-picture of bosonic fractional Chern insulators at high filling factors in multiple flat-band systems}

\author{Licheng Wang}
\affiliation{National Laboratory of Solid State Microstructure, Department of Physics, Nanjing University, Nanjing 210093, China}

\author{Dong-Hao Guan}
\affiliation{National Laboratory of Solid State Microstructure, Department of Physics, Nanjing University, Nanjing 210093, China}

\author{Ai-Lei He}
\email{heailei@yzu.edu.cn}
\affiliation{College of Physics Science and Technology, Yangzhou University, Yangzhou 225002, China}

\author{Shun-Li Yu}%
\email{slyu@nju.edu.cn}
\affiliation{National Laboratory of Solid State Microstructure, Department of Physics, Nanjing University, Nanjing 210093, China}
\affiliation{Jiangsu Key Laboratory of Quantum Information Science and Technology, Nanjing University, Nanjing 210093, China}

\author{Yuan Zhou}%
\email{zhouyuan@nju.edu.cn}
\affiliation{National Laboratory of Solid State Microstructure, Department of Physics, Nanjing University, Nanjing 210093, China}
\affiliation{Jiangsu Key Laboratory of Quantum Information Science and Technology, Nanjing University, Nanjing 210093, China}

\date{\today}

\newcommand*\mycommand[1]{\texttt{\emph{#1}}}
\newcommand{\red}[1]{\textcolor{red}{#1}}
\newcommand{\blue}[1]{\textcolor{blue}{#1}}
\newcommand{\green}[1]{\textcolor{green}{#1}}

\begin{abstract}
Most fractional quantum Hall states have been traditionally identified within a single energy band, such as the lowest Landau level or topological flat band. As more particles are introduced, they inevitably populate higher energy bands. Whether the inclusion of multiple topological bands leads to new physics remains an open question. Here, we propose a universal picture applicable at higher filling factors $\nu
\geq 1$ in bosonic systems: the occupied bands tend to coalesce into an effective single topological band characterized by a total Chern number $\vert C\vert$, the sum of  the Chern number of all occupied lower topological flat bands. Using a Kekul\'{e} lattice model with two lower flat bands featuring a total Chern number $C=1$, regardless of their specific configurations, we identify the emergence of a $\frac{1}{2}$ fractional Chern insulator (FCI) state at integer filling factor $\nu=1$, followed by the Jain sequence states $\frac{2}{3}$ and $\frac{3}{4}$ at filling $\nu=\frac{4}{3}$ and $\frac{6}{4}$. That is a $\nu+\nu$ picture, rather than the generally expected $1+\nu^{\prime}$ picture, where $\nu^{\prime}$ is the permitted FCI filling factor in the single second topological flat band. Our findings deepen the understanding of FCI states and open avenues for discovering exotic fractional topological phases in multiband systems.
\end{abstract}

\maketitle


\emph{Introduction}--
The emergence of topological state which beyond Landau's paradigm has attracted great attentions and become a focal point of research in condensed matter physics over the past two decades~\cite{order1}. The fractional quantum Hall effect, a the representative of topologically ordered states, was first observed in two-dimensional electronic gas system with partially filled Landau level~\cite{fqah} . Its topological order is characterized by ground states with long-range topological entanglement and quasiparticle with fractional charges~\cite{order1,order2,order3,order4,order5}. The Laughlin's trial wave function successfully explained the $\nu=\frac{1}{m}$ fractional quantum Hall effect~\cite{Laughlin1}. Jain proposed the new fractional quantum Hall states with the sequence $\nu=\frac{p}{2mp\pm 1}$ for integer $p$, naturally generalized the Laughlin's states through the composite-fermion approach~\cite{composite-fermion}. Furthermore, wave functions hosting non-Abelian statistics, such as the Read-Rezayi $Z_k$ parafermion states, were established, enriching the family of fractional quantum Hall states~\cite{Moore-Read1,Read-Rezayi1,Read2}. The fractional Chern insulator (FCI) state~\cite{FCI2,FCI3,FCI4,FCI5,FCI6,FCI16,FCI17,FCI20,FCI21,TFB8}, a lattice version of fractional quantum Hall effect in the absence of an external magnetic field, was subsequently proposed in topological flat band (TFB) systems~\cite{TFB1,TFB2,TFB3}, and has recently been observed in rhombohedral multilayer graphene and twisted bilayer transition metal dichalcogenides~\cite{twist_fci1,twist_fci2,twist_fci3,twist_fci4,twist_fci_gra}.

To date, most studies on fractional quantum Hall states have concentrated on the partially filled lowest single band at low filling, either in the Landau level or TFB. This is typically achieved by identifying the bands with large energy gaps or employing projection Hamiltonians that neglect level mixing effect from higher bands. Potential non-Abelian states were reported at higher filling factors, such as the Pfaffian $\nu=\frac{5}{2}$ state~\cite{state-52, Moore-Read1}, the parafermion $\nu=\frac{12}{5}$ state~\cite{state-125}, and the recently proposed $\nu=\frac{3}{2}$ state in twisted bilayer MoTe$_2$~\cite{twist_qsh,non-abelian-twist2}. Nevertheless, the main physics of these states still originates from a single band (the second topological band) with fully occupied the lowest band, i.e., a $1+\nu^{\prime}$ FCI picture with $\nu^{\prime}$ the allowed FCI filling factor on the single higher Landau level or TFB. In fact, the TFB systems can host the multiband characteristic, such as the recently reported second moir\'{e} flat band in twisted bilayer MoTe$_{2}$~\cite{twist-np-higher-TFB1,twist-np-higher-TFB2}. At higher filling factors, the loaded particles inevitably occupy the multiple bands. This raises the fundamental questions: \textit{Can the proposed FCI states on a single TFB be observed in real system including multiple bands, and will the inclusion of multiple bands introduce new physics beyond the single-band picture?} Indeed, a recently numerical study on moir\'{e} FCIs in rhombohedral graphene showed that the $\nu=\frac{1}{3}$ FCI observed in one-band exact diagonalization (ED) algorithm collapses due to band mixing~\cite{moire-ed1}.

In this paper, we employ real-space ED without any projection and the infinite density matrix renormalization group (iDMRG) algorithm to explore exotic FCI states of hard-core bosons at higher filling $\nu\geq 1$ on Kekul\'{e} lattice model with two lower TFBs. We numerically identify a universal $\frac{1}{2}$ FCI state emerging at $\nu=1$ when the two lower TFBs collectively contribute a total Chern number $\vert C_{\text{tot}}\vert=1$, regardless of the topological details of the constituent bands, even when including topological trivial bands. The subsequent $\frac{2}{3}$ and $\frac{3}{4}$ Jain sequence states at filling $\nu=\frac{4}{3}$ and $\nu=\frac{6}{4}$ are also verified. Those findings demonstrate a distinct picture in bosonic multi-TFB models at high filling $\nu\geq 1$: the lower TFBs tend to coalesce into an effective single topological band, rather than the generally expected $1+\nu^{\prime}$ picture on higher TFB. Our results offer a flexible approach to search FCI states in multiband systems.

\begin{figure}[!htb]
	\begin{center}
		\includegraphics[width=0.95\linewidth]{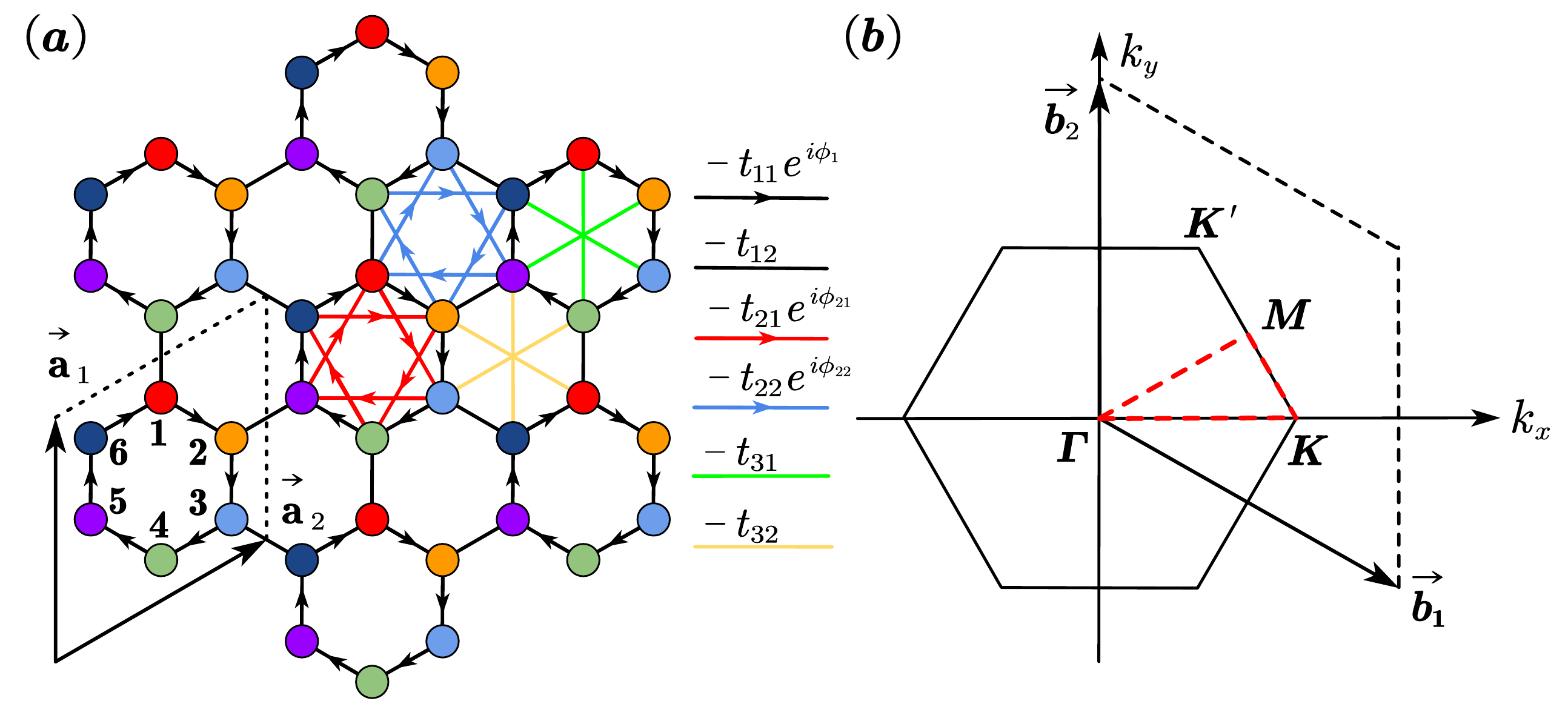}
	\end{center}
	\caption{(a) Schematic structure of the Kekul\'{e} lattice model. $\vec{a}_1$ and $\vec{a}_2$ are lattice vectors, and the six atomic orbitals in the unit cell (dotted rhombus) are labeled from ``1'' to ``6''. Various hopping terms are illustrated by lines with distinct colors (arrows indicate staggered fluxes), and the right panel displays the associated hopping terms. (b) The first Brillouin zone of the Kekul\'{e} lattice. The high-symmetry path $\Gamma$-$K$-$M$-$\Gamma$ are marked by red dash line. }
	\label{Fig1}
\end{figure}

\emph{Model and methods}--We construct a honeycomb lattice with Kekul\'{e}-like hopping terms~\cite{kekule1,kekule2,kekule3}, with six atomic orbitals per unit cell. We consider the first nearest-neighbor(NN1), the second nearest-neighbor(NN2) and the third nearest-neighbor (NN3) hopping terms with staggered fluxes introduced selectively on some of hopping terms. Within the Kekul\'{e} lattice model[Fig~\ref{Fig1}(a)], the hopping terms are classified into two categories, analogous to the intra- and inter-cell hoppings. We introduce staggered fluxes on the NN1 and NN2 hopping terms, while preserving the sixfold rotational symmetry of the system. Consequently, multifold staggered fluxes enable the realization of multiple topological flat bands with distinct Chern number combnations within the Kekul\'{e} lattice model, establishing an ideal platform for searching possible FCI states beyond single band restriction at higher filling.

The noninteracting Hamiltonian is written as
\begin{align}
	H_0 = {} & -t_{11}\sum_{\langle ij \rangle} e^{i\phi_1}b_{i}^{\dagger}b_{j}
	-t_{12}\sum_{\langle ij \rangle} b_{i}^{\dagger}b_{j}  \nonumber \\
	& -t_{21}\sum_{\langle\langle ij\rangle \rangle} e^{i\phi_{21}}b_{i}^{\dagger}b_{j}
	-t_{22}\sum_{\langle\langle ij\rangle \rangle} e^{i\phi_{22}}b_{i}^{\dagger}b_{j}  \nonumber \\
	& -t_{31}\sum_{\langle\langle\langle ij\rangle\rangle \rangle} b_{i}^{\dagger}b_{j}
	-t_{32}\sum_{\langle\langle\langle ij\rangle\rangle \rangle} b_{i}^{\dagger}b_{j} + \text{H.c.}
    \label{eq:H0}
\end{align}
Here $t_{11}$ and $t_{12}$ denote the intra- and inter-cell hopping amplitudes for NN1, $t_{21}$ and $t_{22}$ for NN2, and $t_{31}$ and $t_{32}$ for NN3, respectively. The staggered phase factors $\phi_1$ (on NN1 bonds) , $\phi_{21}$ and $\phi_{22}$ (on NN2 bonds) are engineered to generate the TFBs. The operator $b_{i}^{\dagger}$($b_{i}$) creates (annihilates) a hard-core boson at the $i$th site. To stablize the FCI states, we introduce hard-core bosons with short-range repulsive interactions, The interaction Hamiltonian is given by
\begin{align}
	H_I = V_1 \sum_{\langle ij \rangle} n_i n_j,
\end{align}
with $n_i=b_{i}^{\dagger}b_{i}$ representing the number operator of hard-core bosons at the site $i$, and $V_1$ denoting the NN1 repulsion strength.

We employ ED and iDMRG methods to investigate the topological states and topological orders in this model. In ED calculations, we consider a finite system of $N_x \times N_y$ unit cells (total number of sites $N_s=6 \times N_x \times N_y$), and periodic boundary conditions have been considered. We denote the boson numbers as $N_B$, and the filling factor of the flat band is $\nu=N_B/(N_x \times N_y$). The momentum vector $\textbf{q} = (2 \pi k_1 / N_1, 2 \pi k_2 / N_2)$ is labeled as $(k_1,k_2)$. To study the topological state at larger scales and obtain additional numerical evidence beyond the ED limitations, we also employ the iDMRG algorithm on an $L_x \times L_y$ cylinder geometry with infinite length $L_x \rightarrow \infty$ and finite circumference $L_y$~\cite{dmrg1,dmrg2,dmrg3,dmrg5}.

\emph{Multiple low energy flat bands in Kekul\'{e} lattice}-- The bulk energy spectrum is obtained by numerical diagonalization of the Hamiltonian $H_0$ \hyperref[eq:H0]{Eq.~(1)} following the Fourier transformation. The Chern number of the $n$th band is defined as $C_n = \frac{1}{2\pi} \int_{\mathrm{BZ}} \mathrm{d}^2\mathbf{k}  \mathcal{F}_n(\mathbf{k})$, where $\mathcal{F}_n(\mathbf{k})=\nabla \times \mathcal{A}_n(\mathbf{k}) $ is the Berry curvature, while $\mathcal{A}_n(\mathbf{k})$ denoting the Berry connection. The flatness ratio $f_n$ of the $n$th band is defined as $f_n=\Delta_{n} / W_n $, where $\Delta_{n}$ denotes the band gap between the $n$th and the $(n+1)$th bands, and $W_n$ represents the bandwidth. We explore the parameter space to achieve the multiband exhibiting simultaneously high $f_1$ and $f_2$ values.

\begin{figure}[!htb]
	\begin{center}
		\includegraphics[width=1.0\linewidth]{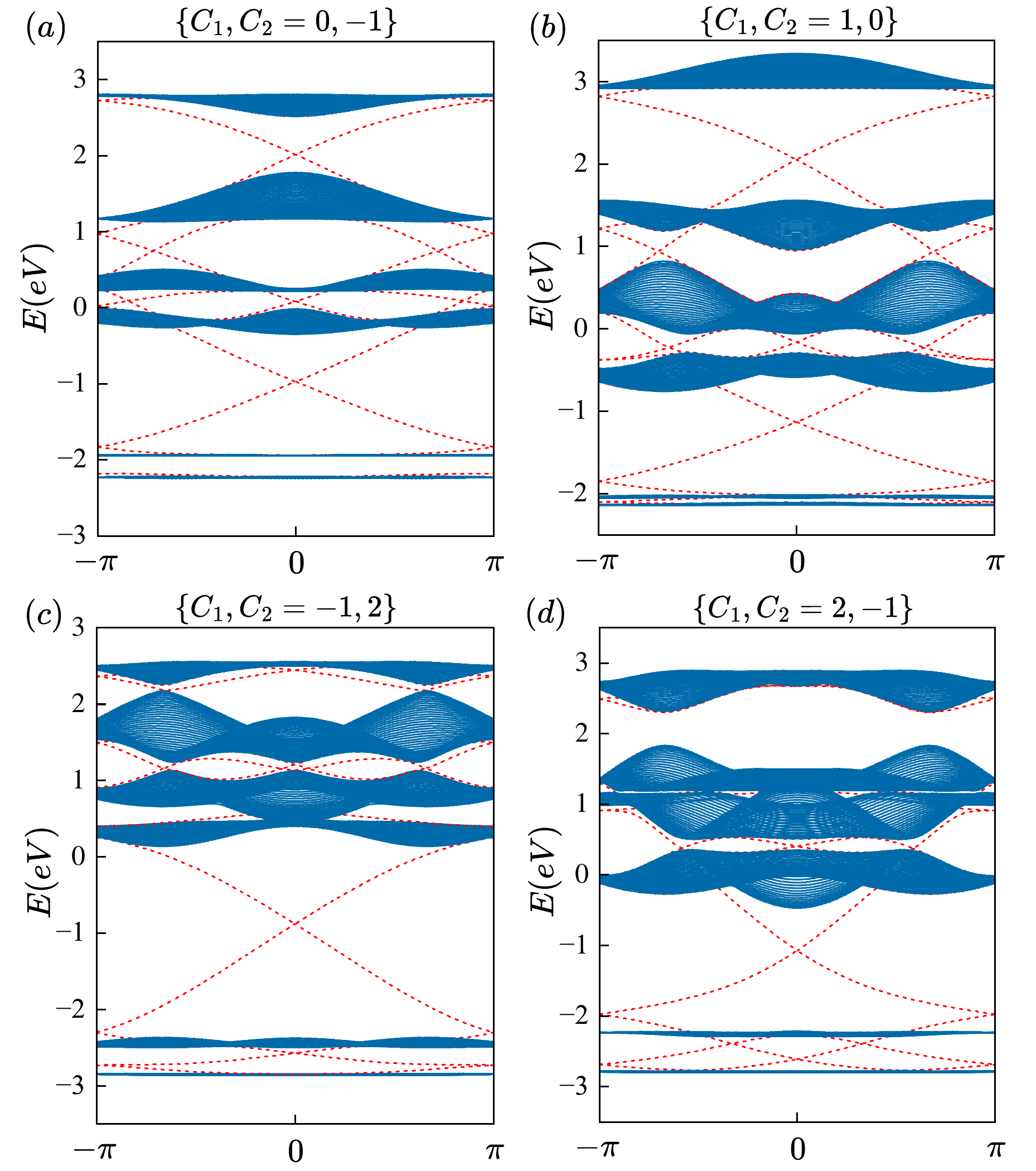}
	\end{center}
	\caption{ The energy spectrum of multiband with distinct Chern number combinations on the cylinder geometry, bulk and edge states are represented by blue solid lines and red dashed lines, respectively. (a) multiband with Chern number $\{C_1,C_2\}=\{0,-1\}$. (b)  $\{C_1,C_2\}=\{1,0\}$. (c) $\{C_1,C_2\}=\{-1,2\}$. (d) $\{C_1,C_2\}=\{2,-1\}$. The detailed parameters are shown Supplementary Materials (SM). }
	\label{Fig2}
\end{figure}

To generate both high-Chern number and the multiple TFBs, the distant-neighbor hopping terms and multifold staggered fluxes are essential~\cite{lattice1,lattice2,TFB8}. When including the hopping terms up to the NN2 with staggered fluxes, the parameter space search over $\{ t_{11},t_{12},t_{21},t_{22},\phi_1, \phi_{21},\phi_{22}\}$ reveals abundant lower double flat bands hosting a total Chern number $|C_{total}|=1$. Most of them exhibit Chern numbers $|C_1|=1$ or $0$ for the lowest flat band. Notably, TFBs with a higher Chern number ($C\geq 2$) are rare and host low flatness ratio. We select three representative lower energy double flat bands cases with distinct Chern number combinations, $\{ C_1,C_2=0,-1\}$ , $\{ C_1,C_2=1,0\}$ and $\{ C_1,C_2=-1,2\}$. As shown in Fig~\ref{Fig2}(a)-(c), the lowest bands host Chern numbers $0$, $1$ and $-1$, with the flatness ratios $\{ f_1,f_2 = 10.7, 97.2\}$, $\{ f_1,f_2 = 5.3, 23.6\}$ and $\{ f_1,f_2 = 24.1, 19.6\}$, respectively. To achieve the double flat bands that the lowest band hosts a higher Chern number, we consider the hopping terms up to the NN3 and select a typical double flat bands combination of $\{ C_1,C_2=2,-1\}$ [Fig~\ref{Fig2}(d)], and the flatness ratios are $\{ f_1,f_2 = 13.5, 25.2\}$.

\emph{FCI states at lower filling factors $\nu<1$}--FCI states at low filling factors $\nu<1$ have been well established on the single TFB in many models, including the bosonic FCIs at $\nu=\frac{r}{r|C|+1}$ and fermionic FCIs at $\nu=\frac{r}{2r|C|+1}$~\cite{FCI2,FCI3,FCI16,FCI17,FCI18,FCI19}. We also show the Laughlin-like FCI states on the lowest TFB in SM, including both $\nu=\frac{1}{2}$ FCI state at $\vert C\vert=1$ and $\nu=\frac{1}{3}$ FCI state at$\vert C\vert=2$. Our results confirm the realization of FCI states on the lowest single TFB.

\begin{figure}[!htb]
	\begin{center}
		\includegraphics[width=1.0\linewidth]{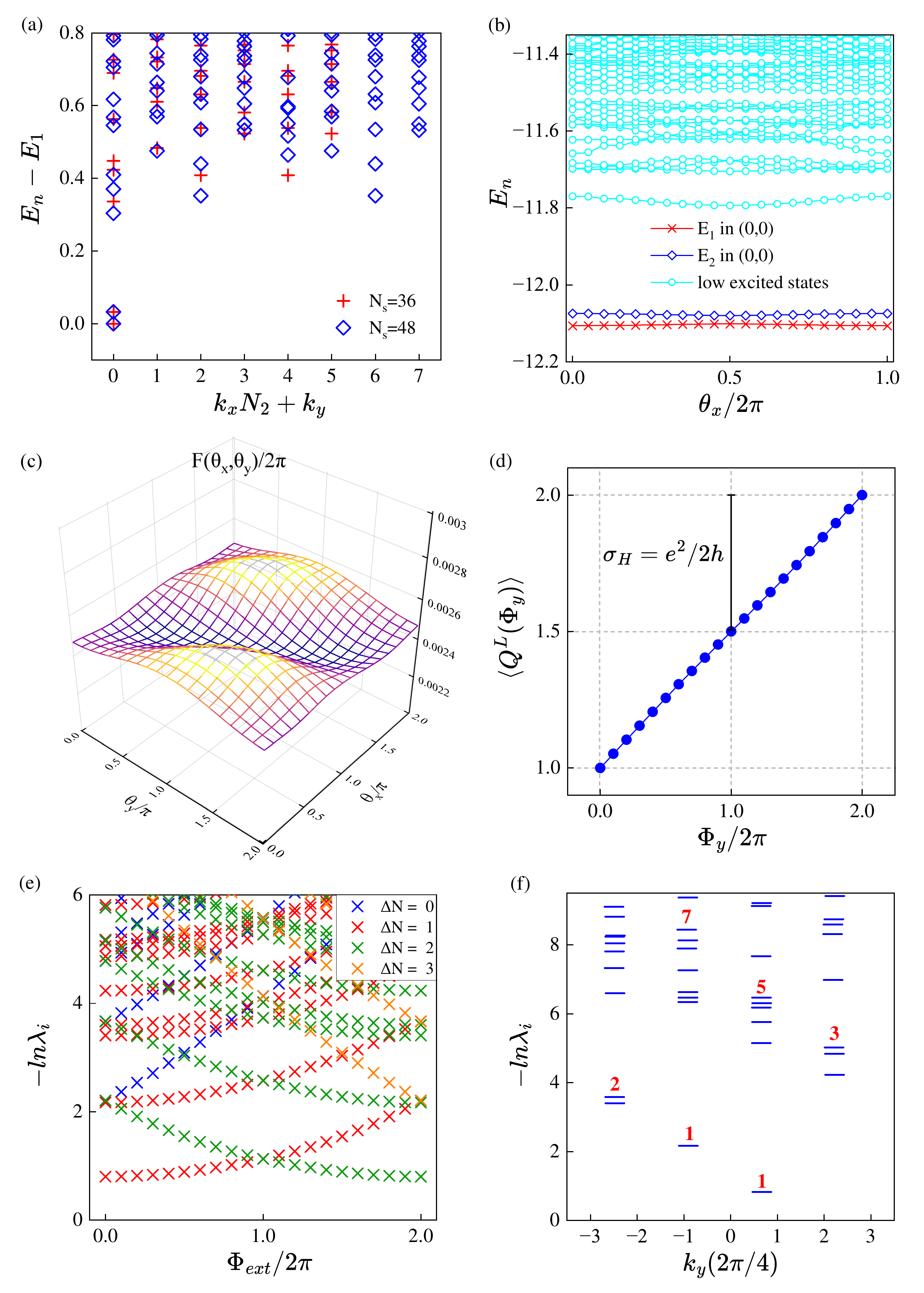}
	\end{center}
	\caption{Topological features at $\nu=1$ for $\{C_1,C_2\}=\{0,-1\}$ combination. (a) Low-energy spectra with $V_1 = 8.0$ for two lattice sizes $N_s =6\times 3\times2=36$ and $N_s=6\times 4\times 2=48$. (b) Low-energy spectrum versus $\theta_x$ at a fixed $\theta_y = 0$ in the $N_s=36$ lattice. (c) Total Berry curvature of the ground states at $20\times 20$ mesh points, indicating a total Chern number $C=1$. (d) Charge pumping after two flux quanta insertion on a cylinder with $L_y=4$ and $\chi=800$. (e) Entanglement spectrum evolution as a function of flux, four charge sectors $\Delta N$ are marked by blue, red, green and orange. (f) Momentum-resolved entanglement spectrum on a cylinder with $L_y=4$ and $\chi=1800$, revealing one chiral edge mode with counting sequence (1,1,2,3,5,7,$\cdots$). }
	\label{Fig3}
\end{figure}

\emph{$\frac{1}{2}$ FCI state at integer filling $\nu=1$}--We start from the integer filling $\nu=1$ at the situation $\{C_1, C_2=0, -1\}$. This case features a topological trivial lowest flat band ($C_1=0$), which usually does not support a FCI state with the single-band picture, and thus provide an ideal platform to check the role of multiple TFBs. The low energy spectra of two lattice sizes $N_s = 36$ and $48$ with $V_1=8.0$ at filling $\nu=1$ [Fig~\ref{Fig3}(a)] exhibit two quasi-degenerate ground states at $(k_x,k_y)=(0,0)$ in both sizes.  For $N_s = 36$, each ground state evolves adiabatically into itself and maintain a robust energy gap from the low excited states under the twist boundary conditions [Fig~\ref{Fig3}(b)]. The many-body Chern number of ground states is defined by $C = \frac{1}{2\pi}\int\int d\theta_x d\theta_y F (\theta_x , \theta_y)$. Here, $F (\theta_x , \theta_y)$ is the many-body Berry curvature $F (\theta_x , \theta_y) = Im(\langle \frac{\partial\Psi}{\partial \theta_y}|  \frac{\partial\Psi}{\partial \theta_x} \rangle - \langle \frac{\partial\Psi}{\partial \theta_x}|  \frac{\partial\Psi}{\partial \theta_y} \rangle)$, and $\Psi$ denotes the wave function of the ground state~\cite{chernnumber-manybody}. Numerical results show that each of these two ground states contributes a Chern number of $\frac{1}{2}$, yielding a total many-body Chern number $|C|=1$. The remarkably smooth Berry curvature of $2\pi$ [Fig~\ref{Fig3}(c)] confirms that the two ground states collectively yield a total Chern number $|C|=1$. Furthermore, we obtain the $\langle Q^L (\Phi_y) \rangle$ [Fig.S3(d) in SM] by iDMRG ($\langle Q^L (\Phi_y) \rangle$ is the charge polarization of left half infinite cylinders when flux $\Phi_y$ threads the cylinder). It is apparent that the wave function pumps $\frac{1}{2}$ charge after a flux quanta threading, corresponding to Hall conductivity $\sigma_H \approx 0.5 e^2 / h$~\cite{dmrg4}. We further calculate the entanglement spectrum as a function of flux $\Phi_y$ [Fig~\ref{Fig3}(e)].
The low-lying eigenvalues of the reduced density matrix in charge sectors $\Delta N=0$ and $\Delta N=2$ are degenerate at $\Phi_y=0$. After $2\pi$ flux insertion, the symmetry center of spectrum undergoes a shift from $\Delta N=1$ to $\Delta N=\frac{3}{2}$, corresponding to a net charge transfer of $\frac{1}{2}$ after inserting a flux quantum. The entanglement spectrum under flux insertion exhibits the characteristics of a gapless edge spectrum. At $\Phi_y=4\pi$, the spectrum becomes symmetric about $\Delta N=2$. Thus, after inserting two flux quanta, the ground state evolves to itself, which is equivalent to the transfer of two anyons from the left edge to the right edge~\cite{dmrg3,dmrg6,dmrg7}.
Moreover, the momentum-resolved entanglement spectrum indicates one branch of the edge modes with counting sequeue $(1,1,2,3,5,7,\cdots)$, consisting with $\frac{1}{2}$ Laughlin state. These features unambiguously identify the emergent $\frac{1}{2}$ Laughlin FCI state at $\nu=1$. Considering the fact that the two lower flat bands host $\{C_1, C_2=0, -1\}$, our numerical results reveal that the loaded hard-core bosons simultaneously occupy the two lower bands, which coalesce into an effective single topological band with $\vert C_{\text{tot}}\vert=1$. The resultant $\frac{1}{2}$ FCI state at integer filling factor is indeed a $\frac{1}{2}+\frac{1}{2}$ state, which cannot be explained in the framework of single-band picture.

Whether such unexpected FCI state is unique for a specific multi-TFB combination. We further study the more double lower energy TFBs while fixing $\vert C_{\text{tot}}\vert=1$ at $\nu=1$, including the combinations $\{C_1, C_2=1,0\}$, $\{C_1, C_2=-1,2\}$ and $\{C_1, C_2=2,-1\}$ as shown in SM. The two quasi-degenerate ground states at fixed momentum point $(k_x,k_y)=(0,0)$ on torus geometry, the low-energy spectrum and many-body Chern numbers (each ground state contributes $C=\frac{1}{2}$), and $\frac{1}{2}$ charge pumping after one flux quantum insertion on cylinder geometry, as well as the evolution of entanglement spectrum as function of threading flux, well support the emergence of $\frac{1}{2}$ Laughlin FCI at integer filling. Our results thus reveals the intrinsic tendency of hard-core bosons to occupy the effective $\vert C_{\text{tot}}\vert =1$ topological flat band, which consists of two low-energy flat bands, leading to exotic topological states beyond the single-flat-band picture. Such combination of multiple bands to form an effectively single TFB was also observed in recently proposed hyperbolic lattice~\cite{hyperbolic1,hyperbolic2,hyperbolic3,hyperbolic4}, indicating the unique and common properties of multi-band systems in hard-core systems. Similar $\frac{1}{2}$ FCI state at $\nu=1$ with Chern band combination $\{C_1, C_2=2,-1\}$ was previously reported on a specific triangular lattice model~\cite{FCI23}. The $\frac{1}{2}$ FCI at $\nu=1$ with Chern band combination $\{C_1, C_2=2,-1\}$ is also in sharp contrast with the numerically proposed bosonic integer quantum Hall state on a generalized Hosftadter lattice with $C=2$ lowest TFB~\cite{biqh2}, where the many-body Chern number is $2$.

\begin{figure}[!htb]
	\begin{center}
		\includegraphics[width=1.0\linewidth]{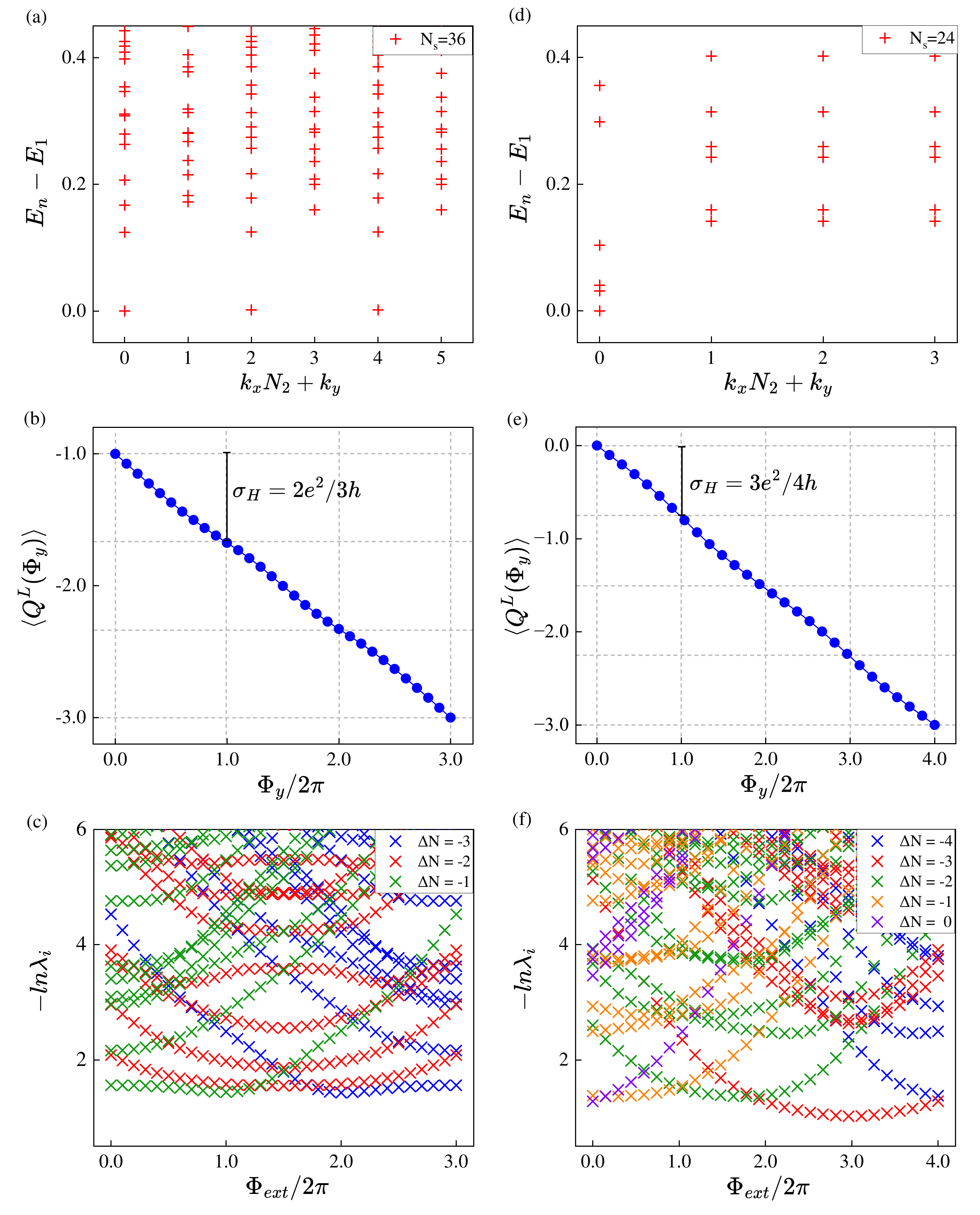}
	\end{center}
	\caption{ Topological features at $\nu=\frac{4}{3}$ with $N_s=6\times3\times 2 =36$(left panels) and at $\nu=\frac{6}{4}$ with $N_s=6\times 2\times 2 =24$ (right panels) for $\{ C_1, C_2 = 1, 0\}$ combination. (a) and (d) Low-energy spectrum; (b) and (e) Charge pumping after flux quanta threading on cylinder geometry with $L_y=3$ and $\chi=600$ ((b)) and $L_y=3$ and $\chi=800$ ((d)); and (c) and (f) Entanglement spectrum evolution as functions of flux insertion.}
	\label{Fig4}
\end{figure}

\textit{Jain sequence states at $\nu>1$}--Having established a new $\nu+\nu$ picture in multi-band system: the loaded hard-core bosons favor to occupy two lower TFBs simultaneously, which generates an effective topological band with $\vert C_{\text{tot}}\vert=\vert C_1+C_2\vert=1$. To verify this picture, we further study the $\nu>1$ cases, in which the loaded hard-core bosons naturally occupy the higher TFB. Whether more FCI states, e.g., the Jain sequence state can be observed in the present multi-TFB systems.

We choose a multi-TFB system with Chern number combination $\{ C_1, C_2 = 1, 0\}$ as example (more in SM). The relatively small band gap between the two flat bands facilitates the realization of hierarchy daughter states~\cite{Hierarchy-Haldane}. The low-energy spectrum for $N_s=36$ at $\nu=\frac{4}{3}$ filling reveals three quasi-degenerate ground states, separated from the low excited states by a gap. The charge pumping after the adiabatic insertion of three flux quanta is $-2$, perfectly matches the quantized Hall conductances of $\frac{2}{3}$ [Fig.~\ref{Fig4}(a) and (b)].
The lowest level of entanglement spectrum [Fig.~\ref{Fig4}(c)] is shifted from $\Delta N=-1$ to $\Delta N=-3$ after inserting three flux quanta, indicating a net charge transfer of $2$ from left edge to the right edge, while the ground states evolving to itself.
 Since the Chern number of the second TFB is $C_2=0$, FCI states are not permitted on the single second TFB band at $\nu= 1+\frac{1}{2}, 1+\frac{2}{3}, \cdots$ with respective two-, three-fold, $\cdots$ ground states degeneracies, where ``1'' denotes the fully occupied lowest band. The observed $\frac{2}{3}$ FCI state at $\nu=\frac{4}{3}$ thus cannot be contributed by the single second TFB band. In contrast, a first daughter state of $\frac{1}{2}$ on an effective single band $\vert C\vert =1$ well matches all the features of topological order. In SM, we show more numerical evidences for the Jain states at $\nu=\frac{4}{3}$. For $C_2=1$, $1+\frac{1}{2}$ (Laughlin-like state, twofold ground state degeneracy with $\nu=\frac{3}{2}$) and $1+\frac{2}{3}$ (first Jain state, threefold ground state degeneracy with $\nu=\frac{5}{3}$) are permitted in $1+\nu^{\prime}$ picture on the second TFB, and thus do not match the observed topological feature at $\nu=\frac{4}{3}$. Similarly, at $\nu=\frac{3}{2}$ $(\frac{6}{4})$, we observe the four-fold quasi-degenerate ground states in the $(k_x,k_y)=(0,0)$ sector, $3$ charge pumping after adiabatic threading four flux quanta [Fig.~\ref{Fig4}(d) and (e)].
The inserting four flux quanta shifts the entanglement spectrum [Fig.~\ref{Fig4}(f)]from $\Delta N=0$ to $\Delta N=-3$, indicating a net charge transfer of $3$ from left edge to the right edge. These topological features manifest the emergent state is a $\frac{3}{4}$ FCI state. Interestingly, the potential $1+\frac{1}{3}$ Laughlin-like FCI state on the second TFB band for $\{ C_1, C_2 = -1, 2\}$ case shares the same similar nature as the first Jain sequence $\frac{2}{3}+\frac{2}{3}$ discussed above. Whether they are exactly the same or distinguishable remains unclear.

We, therefore, identify the general tendency in multi-TFB system--the lower TFBs collectively contribute an effective topological band with $C_{\text{tot}}=\sum_{i\in{\text{occ}}}C_i$-- at high filling factors $\nu\geq 1$. These emergent FCI states show clear differences from the commonly expected $1+\nu^{\prime}$ FCI states in higher TFB. In general, the emergence of FCI requires TFB with high flatness ratio in single TFB systems. It is quite amazing that the robust FCI states can be established at the condition with low flatness, e.g., the flatness ratio of the effective single topological band is as low as $f_{\text{tot}}\sim 4$ (SM). Our results provide new opportunity to search robust FCI states in multi-band system with more flexible requirements, even for the condition of trivial topology.

\emph{Conclusion and discussions}--We reveal a universal picture in strongly correlated multi-TFB systems at high fillings $\nu\geq 1$: the lower TFBs coalesce into an effective topological band, with its Chern number being the sum of the Chern numbers of the occupied lower bands, irrespective of their specific configurations. Based on a Kekul\'{e} lattice model with two lower TFBs and a total Chern number $\vert C_{\text{tot}}\vert=1$, we numerically identify that the emergence of the $\frac{1}{2}$-Laughlin-like bosonic FCI at integer filling $\nu=1$, as well as the subsequent $\frac{2}{3}$ and $\frac{3}{4}$ Jain state at higher filling $\nu=\frac{4}{3}$ and $\nu=\frac{3}{2}(\frac{6}{4})$. Those $\nu+\nu$ FCI states on multi-TFB system break the widely expected $1+\nu^{\prime}$-picture on higher TFB. Our findings highlight the essential difference of fractional quantum Hall states between the single- and multi-TFB system, and reveal unique FCIs which cannot be achieved via single-band projection method.

So far, only bosonic FCI states on the effective $\vert C\vert =1$ condition is well established. Whether the bosonic FCI states can be realized on the effective single bands with higher Chern number, as well as the fermionic counterparts, remain unknown. Moreover, since the hard-core bosons with higher filling factors tend to occupy the multiple bands, whether two distinct FCI states, each of which is realized in the respective TFBs, will emerge under suitable conditions is quite interesting. Systems with multiple TFBs provide the possibility for the realization of such composite FCI states, is an interesting problem that deserves further study. In addition, recently theoretical studies proposed the Moore-Read Pfaffian states in twist transition metal dichalcogenides~\cite{non-abelian-twist1,non-abelian-twist2,non-abelian-twist3}. The single-particle band structure of this system hosts three moir\'{e} flat bands with unit Chern numbers $1$, forming a multiband system. The interplay between multiband occupation and particles with more complex interactions, may yield some novel non-Abelian FCI states. The relative looser condition for the FCI state observed in multiband system may provide alternative access for the non-Abelian states. \\

\emph{Acknowledgement}--This work was supported by the National Natural Science Foundation of China (Grants No. 12204404 No. 12374137, and No. 12434005) and the Natural Science Foundation of Jiangsu Province (Grant No. BK20231397).

\emph{Data availability}--The data that support the findings of this article are not publicly available. The data are available from the authors upon reasonable request.

\bibliography{reference}
\end{document}